# Analysis of microroughness evolution in X-ray astronomical multilayer mirrors by surface topography with the MPES program and by X-ray scattering

R. Canestrari[*a], D. Spiga[a], G. Pareschi[a]

[a]INAF - Osservatorio Astronomico di Brera, Via E. Bianchi 46, Merate (LC), Italy I23807

**ABSTRACT**

Future hard X-ray telescopes (e.g. SIMBOL-X and Constellation-X) will make use of hard X-ray optics with multilayer coatings, with angular resolutions comparable to the achieved ones in the soft X-rays. One of the crucial points in X-ray optics, indeed, is multilayer interfacial microroughness that causes effective area reduction and X-Ray Scattering (XRS). The latter, in particular, is responsible for image quality degradation. Interfacial smoothness deterioration in multilayer deposition processes is commonly observed as a result of substrate profile replication and intrinsic random deposition noise. For this reason, roughness growth should be carefully investigated by surface topographic analysis, X-ray reflectivity and XRS measurements. It is convenient to express the roughness evolution in terms of interface Power Spectral Densities (PSD), that are directly related to XRS and, in turn, in affecting the optic HEW (Half Energy Width). In order to interpret roughness amplification and to help us to predict the imaging performance of hard X-ray optics, we have implemented a well known kinetic continuum equation model in a IDL language program (MPES, Multilayer PSDs Evolution Simulator), allowing us the determination of characteristic growth parameters in multilayer coatings. In this paper we present some results from analysis we performed on several samples coated with hard X-ray multilayers (W/Si, Pt/C, Mo/Si) using different deposition techniques. We show also the XRS predictions resulting from the obtained modelizations, in comparison to the experimental XRS measurements performed at the energy of 8.05 keV.

## 1. INTRODUCTION

In the next years a number of hard X-ray telescopes will fly aboard satellites like SIMBOL-X[1] and Constellation-X[2]. One of the answers to the problem of reflection and focalization of hard X-rays is the use of optics with multilayer reflecting coatings. The main advantage related to the use of multilayers is the great enhancement achievable in reflectivity and effective area of the mirror shell up to 70–80 keV for grazing incident angles of 0.1–0.2 degrees. Indeed, it is well known that the deposition of thin films causes, generally, a degradation of the surface smoothness. This effect is more evident if the number of deposited layers is large, as in the case of multilayer coatings.

The topographic characteristics of a reflecting surface (expressed in terms of microroughness Power Spectral Density - PSD) can be related, through the perturbation theory[3], to the amount of radiation scattered by the mirror in the reflection process: consequently, angular resolution performances (usually given in terms of HEW, *Half-Energy-Width*) of an X-ray mirror shell will be necessarily affected by the smoothness properties of mirror surface. A similar relationship can be stated, for a multilayer-coated surface, between the X-ray scattering and the roughness profiles of multilayer interfaces. The interface roughening in deposition of multilayer coatings on a mirror can be regarded as a layer-by-layer amplification of the microroughness Power Spectral Density (PSD) of each deposited layer, occurring mainly in the spatial wavelength range [1÷0.05] μm. In the following, we will refer this spatial wavelength amplification as a microroughness growth/evolution. It should be reminded that the rms microroughness rms σ in a given frequency interval [$f_{min} \div f_{max}$] is related to the surface PSD as

$$\sigma = \left( \int_{f_{min}}^{f_{max}} PSD(f)\,df \right)^{1/2} \tag{1}$$

[*] rodolfo.canestrari@brera.inaf.it phone +39 039 9991104; fax +39 039 9991160

moreover, it is well known that the reflectivity of an X-ray mirror with surface rms decays exponentially with $\sigma^2$ by X-ray scattering[4]:

$$R = R_F \exp\left(-\frac{16\pi^2 \sin^2\theta_i \sigma^2}{\lambda^2}\right) \qquad (2)$$

therefore, the effect is more severe for harder X-rays. The X-rays are scattered in the surrounding directions according to the PSD trend[3], hence a mirror surface PSD measurement over a *very* wide spatial wavelength scan makes possible to predict the X-ray scattering from the mirror surface: this determines, in addition to the energy-independent HEW term due to mirror shape deformations, the degradation at high energies of the X-ray focusing mirror angular resolution. We provided in fig. 1 an example of HEW simulation as a function of X-ray photon energy, for the *simplified* case of a single-layer with a surface characterized by a PSD analyzed in this work. However, the case of a hard X-rays optic with multilayer coating is more complicated, because of the multiple reflections/scattering in the stack: it is now easy to understand the importance of an accurate investigation of the adopted substrate finishing characteristics and microroughness evolution in predicting the imaging performance of a hard X-ray optic. Finally, the PSD evolution analysis can also cast light on possible improvements of the adopted deposition process.

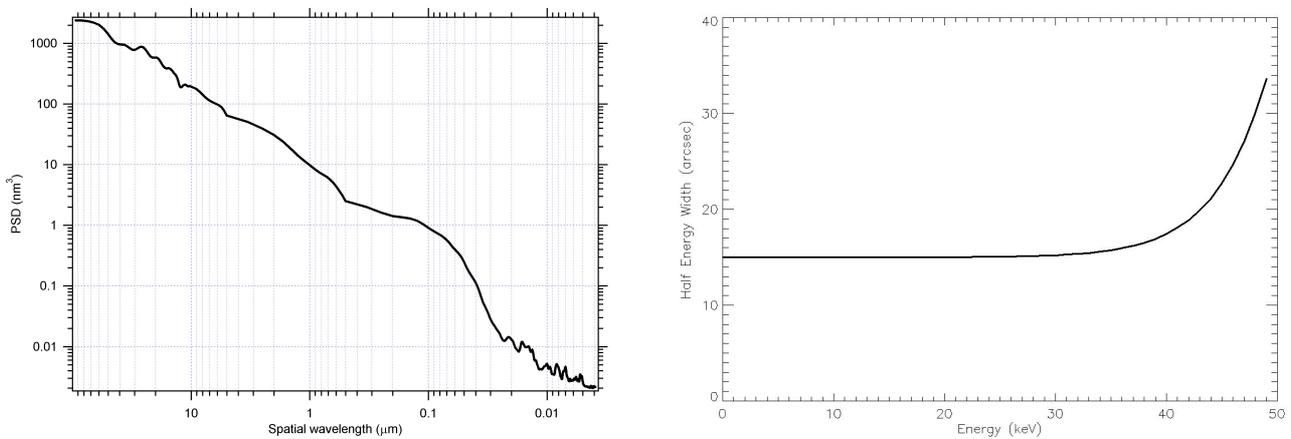

Figure1: (left) PSD of the outer surface of a Mo/Si multilayer deposited at *Laboratori Nazionali di Legnaro, INFN, Italy* (deposited with the same technique of the sample D). The substrate is a single Au layer replicated from a superpolished fused silica sample by Ni electroforming at *Media-Lario techn (Italy)*.
(right) *Simulated* HEW as a function of the photon energy for a *single-layer* X-ray optic with a surface microroughness characterized by the PSD on the left-hand graph. The grazing incidence angle is 0.18 deg (a typical SIMBOL-X[1] one): the results are added in quadrature to 15 arcsec of HEW, ascribed to mirror shape errors. In order to compute the HEW down to the low energies, the PSD has been extrapolated at low spatial frequencies, assuming a typical power-law spectrum (the spectral index is near 1.5).

As the surface topography of inner layers in a multilayer is not directly observable, the interfacial PSD evolution has to be inferred *indirectly* from the PSD of the outermost layer. To study and interpret the roughness evolution in terms of layer PSD, we made use of a well known kinetic continuum equation model[5,6], already applied in the past by Stearns[7] (and successively by Spiller[8]) to describe the PSD evolution in a multilayer coating. With this approach, the surface PSD of a deposited layer is conceived as the sum of two distinct contributions:
  a) the surface roughening intrinsically related only to the deposition process properties and to the layer thickness, associated to an *intrinsic PSD*,
  b) a partial replication of the underlying layer topography. For the first deposited layer the substrate profile is partially replicated.

In this model, the microroughness evolution along the layers stack is driven by few *growth parameters*. In order to infer typical values of these parameters for different deposition techniques and, consequently, to investigate the microroughness evolution in multilayers for future hard X-ray optics, the adopted model has been implemented in the IDL-based program *Multilayer PSDs Evolution Simulator* (MPES). The program was developed by one of us (R. Canestrari) and will be made available online in the next future.

The MPES program has been validated for several multilayer coating (W/Si, Mo/Si) samples, fabricated using different deposition techniques. As input data we used the PSD characterization in a wide spectral range [200–0.02] μm of the substrate before deposition, and the PSD of the multilayer outer surface. The PSD data are obtained from the superposition of results of several techniques, like Atomic Force Microscopy (AFM), optical profilometry (WYKO) and X-ray scattering (only for substrate). Since the PSD evolution is strongly affected by the thickness of layers, we used the values derived from analysis of X-Ray Reflectivity (XRR) measurements. Starting from these known data and from initial growth parameters values, the growth model is applied obtaining the expected multilayer surface PSD. The parameter values are then manually adjusted until a satisfactory agreement is reached between the modeled and the experimental outer PSD. The parameters found from this analysis are then used to compute the internal PSD of the multilayer[7], therefore the PSD evolution can be traced over all the spectral range under analysis. Moreover, we will see that *Cross-Correlations* of couples of boundaries in the multilayer can be recovered also as a function of spatial frequency. The evolution of the internal PSD and the Cross-Correlations allow us to compute the scattering diagram for X-rays when they impinge on the sample e.g. at the 1st Bragg peak incidence angle.

In this work will show the MPES results of the PSD fitting for several multilayer samples. In Sec. 2 we will resume the adopted model[5,6,7,8] for microroughness evolution in thin films. In Sec. 3 we will provide a description of analyzed samples properties, while in Sec. 4 after a short description of the adopted experimental methods we will expose the PSD measurements and the microroughness growth analyses results obtained with MPES. Finally, in Sec.5 we will show an independent check of the achieved results by X-ray scattering, followed by a short discussion.

## 2. MICROROUGHNESS GROWTH MODEL

The surface profile evolution in time $z(x,t)$ can be described through the *Edward-Wilkinson* equation[5]: in the following we will assume that layers are grown at constant rate, therefore we replace the time variable $t$ with the layer thickness $\tau$. We restrict to the simplest form of the evolution equation for $z(x,\tau)$, at the first order[6]:

$$\frac{\partial z(x)}{\partial \tau} = -\nu \left| \nabla^n z(x) \right| + \frac{\partial \eta}{\partial \tau} \tag{3}$$

in eq. (3), the first term accounts for profile smoothing effect caused by surface relaxation processes, while the second one accounts for roughness increase. Here $\nu$ is a proportionality constant related to the intensity of the smoothing process and $\eta$ is a random shot noise term, typical of the used deposition process. A solution for the linear equation (3) can be expressed in terms of the intrinsical layer surface PSD, as suggested by Stearns[7]:

$$P^{\text{int}}(f) = \Omega \frac{1 - \exp\left(-2\nu |2\pi f|^n \tau\right)}{2\nu |2\pi f|^n} \tag{4}$$

that represents the bi-dimensional PSD (i.e., as a function of two spatial frequencies $f_x$, $f_y$: as the sample is supposed to be isotropic, all 2D PSD information is enclosed in the section $P(f_x, 0) = P(f)$ ) of the surface of the deposited layer as a function of the spatial frequency $f$. In equation (3) three growth parameters characterizing the growing layer surface can be identified: $\Omega$ represents the volume of the particles (atoms, nanocrystals) which constitute the growing film, $\nu$ is the coefficient that appears in eq. (3) and $n$ is an integer number related to the slope of the high spatial frequency trend of the PSD. An example of $P^{int}$ functions for different $\tau$ values is plotted in fig. 2. Two different regimes can be recognized: a low spatial frequency domain with a plateau typical of random deposition process (white noise), and a high spatial frequency domain where the PSD trend is a power law (i.e. proportional to $f^{-n}$). The spatial wavelength $l^* = (\nu\tau)^{1/n}$ marks the regime transition. Surface structures with lateral size larger than $l^*$ are enhanced, they are damped out if smaller than $l^*$.

A multilayer coating deposition is the iterative deposition of a number of bi-layers, i.e. layer couples of materials with a different density. The heavier material is called "absorber" and the lighter "spacer". The surface profile of each layer will be the result of the combination of the intrinsic contribution of the deposition process and the partial replication of the underlying layer profile. Following Stearns[5], the layer-by-layer growth is expressed by the linear, iterative equation:

$$P_j(f) = P_j^{\text{int}}(f) + a_j(f) P_{j-1}(f) \tag{5}$$

where $P_j^{int}(f)$ is the intrinsic layer PSD (eq. 4), and $a_j(f)$ is the *replication factor* of $(j-1)^{th}$ layer[4]:

$$a_j(f) = \exp\left(-\nu_j |2\pi f|^n \tau_j\right) \qquad (6)$$

the replication factor is close to 1 with a sudden cutoff at $f = 1/l^*$; at low frequencies of the previous deposited layer is entirely replicated, at higher frequencies its relief is cancelled in favor of the intrinsic PSD term $P_j^{int}(f)$. Iterating equation (5) up to the $N^{th}$ layer the evolution of the 2D PSD from the substrate surface to the outermost layer of the multilayer reflecting coating can be obtained. The rms evolution is computed by integration over the frequencies domain.

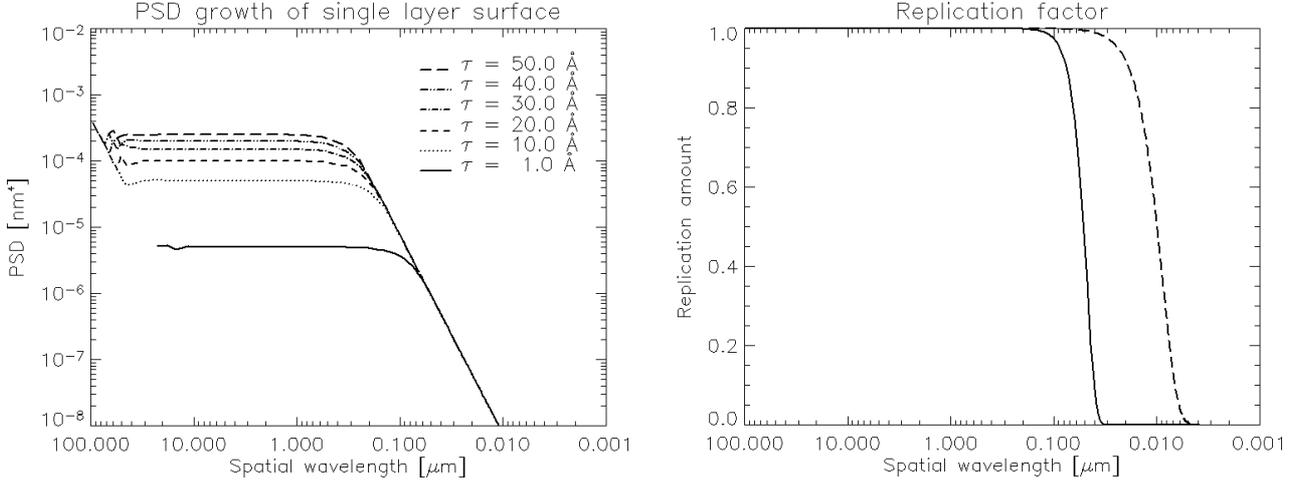

Figure 2: (left) Evolution of the theoretical surface PSD described by the equation (4) when increasing the deposited layer thickness. The growth parameters are $\Omega = 0.05$ nm$^3$, $\nu = 0.0133$ μm$^2$ and $n = 3$.
(right) Examples of replication factors in a multilayer. High-Density material (solid line) and Low-Density (dashed line) material ($\Omega_{H-D} = 0.05$ nm$^3$, $\nu_{H-D} = 1.6 \cdot 10^{-8}$ μm$^4$ and $n_{H-D} = 5$; $\Omega_{L-D} = 0.02$ nm$^3$, $\nu_{L-D} = 9 \cdot 10^{-7}$ μm$^2$ and $n_{L-D} = 3$).

## 3. CHOICE AND DESCRIPTION OF ANALYZED SAMPLES

We listed in tab. 1 the analyzed samples and their properties: all analyzed multilayer samples have an almost constant d-spacing. Even though the exposed model is applicable to graded multilayers also, this choice allows us to simplify the search for growth parameters and speeds up the computation time.

W/Si multilayers are interesting for their astronomical applications in future X-ray missions like SIMBOL-X, Constellation-X and HEXIT-SAT. Mo/Si multilayer mirrors are typically used in EUV nanolithography of electronic components.

The e-beam evaporation coating facility, used to deposit the samples A and B, is the same used to deposit the SAX[9], XMM[10] and SWIFT-XRT[11] soft X-ray optics with single Au layer. This method takes the advantage of a large surface coverage with an high deposition rate. Another advantage of e-beam evaporation is the uniform coating over large surfaces, like mandrels used for mirror shell replication. The main drawback of this method is the evaporation rate instability; very good results were, indeed, obtained in terms of peak reflectivity[12]. Sputtering processes are suitable to deposit compact and uniform films with good optical and mechanical properties: multilayers deposited by sputtering (like sample C) exhibit a good smoothness and durability in time, an important requirement for space telescopes with long lifetime. The deposition rate is, indeed, quite low at the expense of the deposition time.

The multilayer coating of the sample D was deposited using a RF magnetron sputtering facility[13] installed to *Laboratori Nazionali di Legnaro* of the *Istituto Nazionale di Fisica Nucleare* (Italy). The sputtering source is driven by a RF power supply and combines an ion bombardment with a low-voltage substrate bias. As we will show in Sec. 4, these two effects promote the mobility of deposited particles, reducing surface microroughness growth. Hydrogen incorporation in Si layers is also observed: this reduces the Si density, therefore it reduces the X-ray absorption of each bi-layer with the effect of a gain in the multilayer reflectivity[14].

The sample E is a multilayer deposited on a superpolished fused silica substrate with a DC magnetron sputtering facility[15] installed at the Smithsonian Astrophysical Observatory (SAO): this facility is specifically conceived to deposit multilayer coatings on mirror shells preformed by Ni electroforming. The research is aimed to manufacture the hard X-ray optics of Constellation-X[16]. The sample was coated in parallel to a mirror shell in order to separate the intrinsic roughness developed in the sputtering process from the roughness due to the substrate topography replication.

The substrates used for the multilayers growth were all characterized by a good surface quality[17]. Silicon wafers (samples A, B, C) have a typical roughness rms of 3-4 Å. Electroformed Ni substrates (sample D) replicated from fused silica samples are suitable to test multilayer deposition processes for X-ray telescopes, since these substrates are produced with the same manufacturing process proposed for the optics of SIMBOL-X, HEXIT-SAT and Con-X. The fused silica substrate ($\sigma < 2$ Å) used for sample E is a reference substrate for surface smoothness.

Table 1: main features of the analyzed multilayer samples. Multiple values indicate a layer thickness variation through the stack.

| Sample ID | Substrate | Number of bilayers | Absorber | Spacer | Recipe | | Deposition method |
|---|---|---|---|---|---|---|---|
| A | Si wafer | 40 | W | Si | d = 43 Å | $\Gamma$ = 0.37 | e-beam evaporation |
| B | Si wafer | 10<br>10<br>10 | W | Si | d = 58.27 Å<br>d = 56.54 Å<br>d = 52.96 Å | $\Gamma$ = 0.41<br>$\Gamma$ = 0.44<br>$\Gamma$ = 0.34 | e-beam evaporation with ion etching |
| C | Si wafer | 40 | W | Si | d = 54 Å | $\Gamma$ = 0.13 | DC magnetron sputtering |
| D | Ni replicated | 40.5 | Mo | Si | d = 72.7 Å | $\Gamma$ = 0.44 | RF magnetron sputtering |
| E | Fused silica | 7<br>40 | W | Si | d = 130 Å<br>d = 38 Å | $\Gamma$ = 0.355<br>$\Gamma$ = 0.47 | DC magnetron sputtering |

### 4. MEASUREMENT RESULTS AND MICROROUGHNESS GROWTH ANALYSIS

All presented multilayer mirror samples and the corresponding substrates have been widely characterized in topographical properties. The measured PSD are the superposition of single PSD data computed from profiles measured with the optical profilometer WYKO and those calculated from the 100 μm, 10 μm and 1 μm - sized AFM maps. The latter allows the coverage of wavelength band 100÷0.01 μm, the former is sensitive to long wavelengths range 300÷3μm. The topographic measurements were repeated on several points of the samples surface to rule out local features, and the single PSD extracted have been averaged in order to return a statistically significant surface description. Along with X-ray reflectivity tests at 8.05 keV (Cu-K$\alpha$ line) performed with a Bede-D1 diffractometer we have inferred the multilayers structure. In order to interpret these data we have used the IMD[18] program and, for multilayers exhibiting evidence of d-spacing drift along the stack, the PPM[19] program (see also this conference[20]). All multilayers have a periodic or quasi-periodic structure. The thickness values have been used as parameters for the PSD fit with MPES. In the following tables we present the PSD fit results giving the values of $l^*$ instead of $\nu$, and $\Omega/\Omega_0$ rather than $\Omega$ itself, ($\Omega_0$ is the atomic volume of the element, see tab. 2). Since the PSD provided by the evolution model are 2D PSD, we converted all internal PSD to 1D PSD in order to compare them with experimental data. The 1D-2D PSD conversion formulae for isotropic samples are provided in literature[21].

Table 2: Reference values of the atomic volumes adopted for each material.

| | Si | Mo | W |
|---|---|---|---|
| $\Omega_0$ [nm$^3$] | 0.02 | 0.016 | 0.016 |

**Sample A**

Tab. 3 summarizes the measured microroughness values for the substrate ($\sigma_{sub}$) and multilayer ($\sigma_{ML}$) samples. Despite the high roughness, we present anyway the result of this multilayer sample because it exhibits a PSD characterized by a clear deviation from the power-law (with a spectral index near 1.7) trend of the substrate.

We show the measured PSDs of the substrate and the multilayer in fig. 3 (left), including all the intermediate PSDs of the upper face of all 40 W layers. The growth of the PSD, as well as the wide bump around 0.2 μm, is fitted accurately. The PSD growth is mainly localized in the wavelength range 10 ÷ 0.03 μm. The roughness rms evolves accordingly from 5.3 Å to 8.5 Å, over all frequency range. The final $\sigma_{rms}$ value is in good agreement with the roughness rms experimental value and the inferred one from the XRR fit (tab. 3). The $\sigma_{rms}$ increase in the stack is apparent in fig. 3 (right).

The growth parameters inferred with this multilayer film are listed in tab. 4. The higher values of the *n* parameters, with respect to the substrate, indicates a steep cutoff of PSD at high frequencies. This behavior has been already noticed in previous works[7,8] and explained as local relaxation of growing surface. In fact, a high-frequency trend extrapolation of $PSD_{ML}$ would cross the substrate PSD at a frequency near $1/l^*$. Unfortunately, at these frequency the AFM sensitivity does not allow us to validate the extrapolation. The quite high value required for $\Omega_W$ (~25 nm$^3$) suggests the presence of crystallites in the W layers. On the other hand, the $\Omega_{Si}$ parameter equals the atomic volume of Silicon, therefore Si layers should be amorphous. The resulting smoothing effect of Si layers partially compensates the roughening of W layers, causing the σ evolution curve to have a saw-teeth shape superimposed to the overall increasing trend (see fig. 3, right).

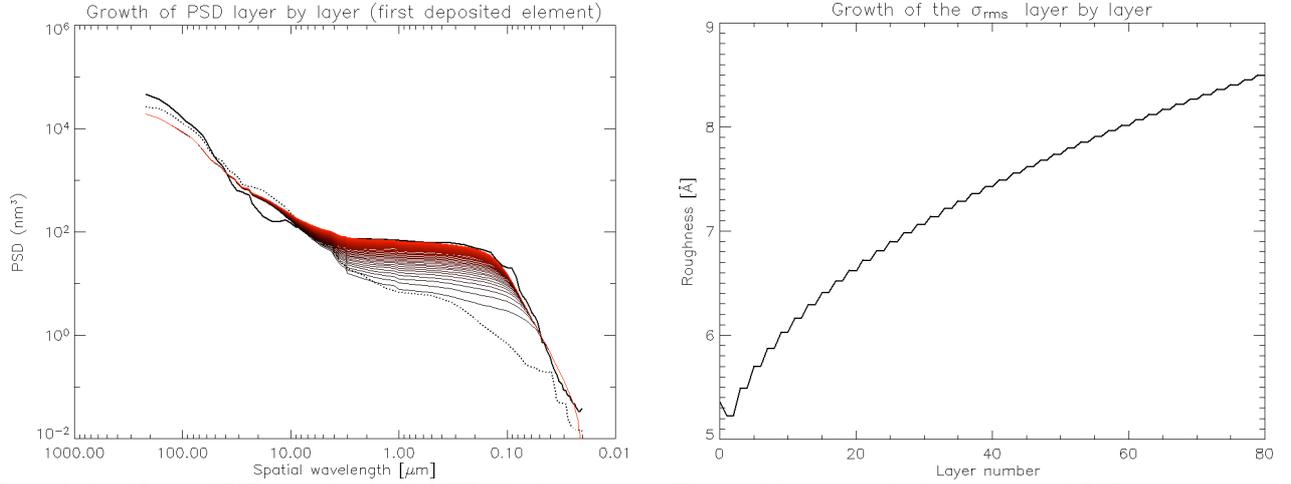

Figure 3: (left) Internal PSDs simulated with MPES for the sample A. The trend fits accurately the experimental PSD.
(right) Microroughness evolution for the multilayer sample A calculated from the internal PSDs. The final rms value is close to 8.5 Å, in good agreement with the $σ_{rms}$ value measured from topography and with the inferred one from XRR.

Table 3: Experimental roughness values for the Si wafer substrate and for the outer surface of the sample A.

| $σ_{sub}$ [Å] | $σ_{ML}$ [Å] | $σ_{XRR}$ [Å] |
|---|---|---|
| 5.1 | 9.16 | 8.2 |

Table 4: Growth parameters used to simulate the microroughness evolution of the multilayer sample A.

|  | Ω | $l^*$ [nm] | n |
|---|---|---|---|
| W layers | 1560 · $Ω_{0,W}$ | 7.6 | 5 |
| Si layers | 1 · $Ω_{0,Si}$ | 1.35 | 3 |

**Sample B**

This 30 bi-layers W/Si multilayer coating has been deposited on a Si wafer substrate using the same e-beam deposition facility of the sample A. In addition, an ion (Ar$^+$) etching source was used to reduce the roughness of W layers.

The effects of ion etching are apparent in fig. 4. The PSD growth has been limited in amplitude by an order of magnitude. The bump frequencies extension also diminished with respect to the sample A. However, a relevant PSD growth appeared in the middle-low frequencies range (200 >$l$ >20 μm), and this brought the final microroughness level ($σ_{ML}$ = 8.7 Å, see tab. 5) near the σ value of sample A.

The PSD evolution obtained from MPES simulation is shown in fig. 4 (left): the fit is satisfactory only in the bump region, i.e. for wavelengths smaller than 20 μm. The layer-by-layer trend of roughness σ for the sample B is also presented in fig. 4 (right). Since the simulated PSD evolution underestimates the low-frequencies contribution, the final σ value computed from simulation (6 Å) is lower than the measured one over the whole wavelength range (8.7 Å, see tab. 5). The parameters used to simulate this growth of the PSD are listed in tab. 6. For instance, the Ω volume for Si equals the atomic volume ($Ω_{Si} = Ω_{0,Si} = 0.02$ nm$^3$), indicating an amorphous growth of Si layers. W layers consist of particles with a typical size of 4-5 times the radius of a W atom ($Ω_W$ ~1.6 nm$^3$). Therefore, the ion etching acted by reducing the size of nanocrystals which W layers are made of, and by promoting the smoothing effect of Si layers. This also results from the increase of $l^*$ parameter with respect to the sample A (comp. tables 4 and 6).

Both simulated and experimental σ values are in disagreement with the inferred roughness value from XRR (12 Å). This mismatch suggests the presence of layers interdiffusion, that degrades the multilayer reflectivity, without being detected from surface profiles; therefore, it cannot be evaluated with MPES.

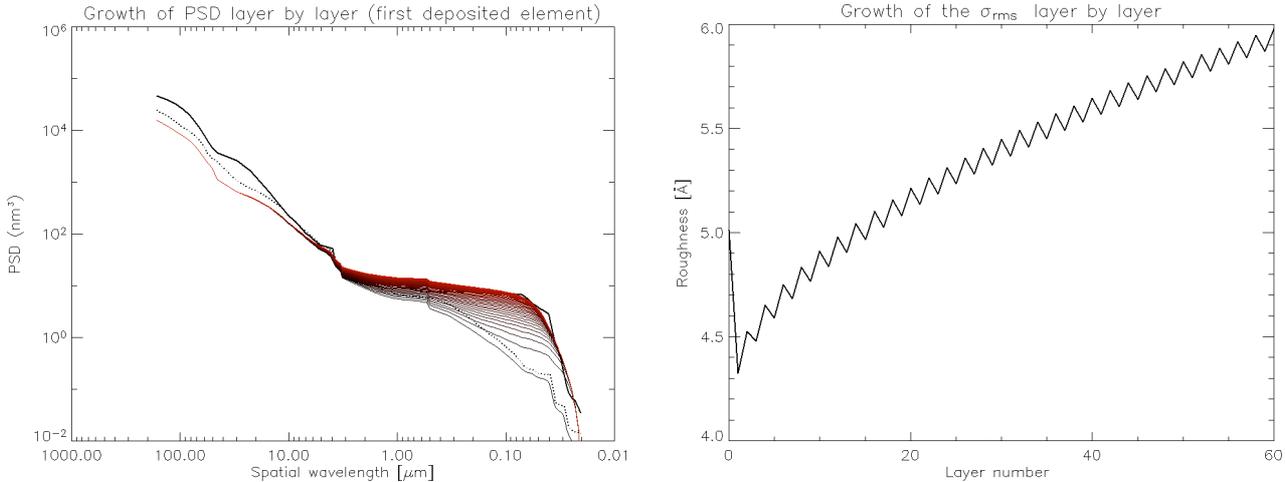

Figure 4: (left) Internal PSDs simulated with MPES for the multilayer sample B. The simulation fits the experimental PSD in the bump area.
(right) Microroughness evolution for the multilayer sample B calculated from the internal PSDs. The "saw-teeth" trend is sharper than in sample A (see fig. 2) due to the intense smoothing action of Si layers.

Table 5: Experimental roughness values for the Silicon wafer substrate and for the external surface of the sample B.

| $\sigma_{sub}$ [Å] | $\sigma_{ML}$ [Å] | $\sigma_{XRR}$ [Å] |
|---|---|---|
| 5.1 | 8.7 | 11.9 |

Table 6: parameters used to simulate the microroughness growth in the multilayer sample B.

|  | $\Omega$ | $l^*$ [nm] | n |
|---|---|---|---|
| W layers | $100 \cdot \Omega_{0,W}$ | 3.3 | 6 |
| Si layers | $1 \cdot \Omega_{0,Si}$ | 3.2 | 5 |

**Sample C**

This multilayer sample was deposited by a magnetron sputtering facility; it exhibits a limited PSD growth, and consequently a low final roughness rms (see tab. 7). The amplitude of the PSD bump is reduced by a factor two with respect to sample B. The result of simulation is shown in fig. 5 (left): the agreement is not perfect, but the final microroughness value is in good agreement with the experimental value $\sigma_{ML}$ (see fig. 5, right, and tab. 7). In this case the PSD was measured only in the spectral range 10 – 0.02 μm.

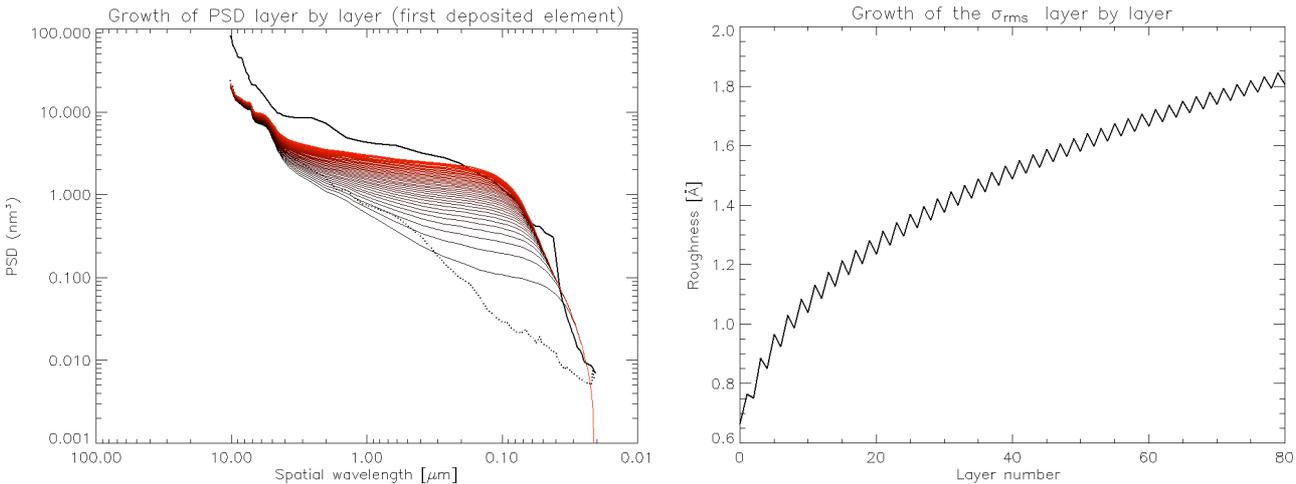

Figure 5: (left) Internal PSDs simulated with MPES for the multilayer sample C.
(right) Microroughness evolution for the multilayer sample C calculated from the internal PSDs.

Table 7: Measured roughness values in the narrow investigated range of integration 10-0.02 μm.

| $\sigma_{sub}$ [Å] | $\sigma_{ML}$ [Å] | $\sigma_{XRR}$ [Å] |
|---|---|---|
| 0.7 | 1.9 | 5 |

Table 8: Growth parameters used to simulate the measured level of microroughness for the external surface of sample C.

|  | $\Omega$ | $l^*$ [nm] | n |
|---|---|---|---|
| W layers | $60 \cdot \Omega_{0,W}$ | 4.4 | 6 |
| Si layers | $4 \cdot \Omega_{0,Si}$ | 4.9 | 5 |

The growth parameters used to simulate this observed trend of the PSD curve are reported in tab. 8: the volume parameter $\Omega$ of W is smaller than for samples A and B, for Si it is only few times the atomic volume. The power-law index *n* have a considerably large value, indicating an intense smoothing effect in the high frequencies range. Interestingly, the values of correlation lengths $l^*_W$ and $l^*_{Si}$ are very similar in this case.

The final value of multilayer rms roughness is much lower than $\sigma_{XRR}$, the roughness rms inferred from X-ray reflectivity (see tab. 7). The discrepancy can be due to the limited interval of investigated wavelengths.

**Sample D**

This Mo/Si multilayer, deposited by RF magnetron sputtering on a replicated substrate by Nickel electroforming from a superpolished fused silica sample, exhibits a different shape from those of samples A, B, C. The surface PSD is visible in fig. 6 (left, black continuum line), mainly under the substrate PSD (dotted line). The final PSD has *decreased* with respect to the substrate in the wavelength range (1 – 0.01 μm), whereas at larger wavelength the surface quality remained unchanged. In other words, the multilayer deposition apparently improved the surface smoothness through the layer surface relaxing, which resulted in a measurable damping of the microroughness from 5.4 (substrate) and 4.8 Å (multilayer surface). The decrease of roughness rms across the deposition can be seen in fig. 6 (right). It worth noticing that most of surface smoothing action is exerted by Si layers, even though the relatively small size of Mo grains also contributes to the PSD decay.

The smoothing of microroughness at high frequencies, that reduces the large-angle scattering and improves the optical performances in hard X-rays, makes of the adopted deposition method a very good candidate to deposit multilayers for hard X-ray astronomical optics.

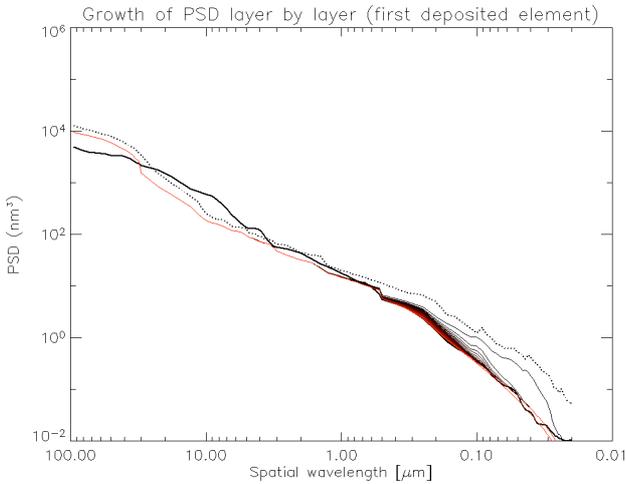 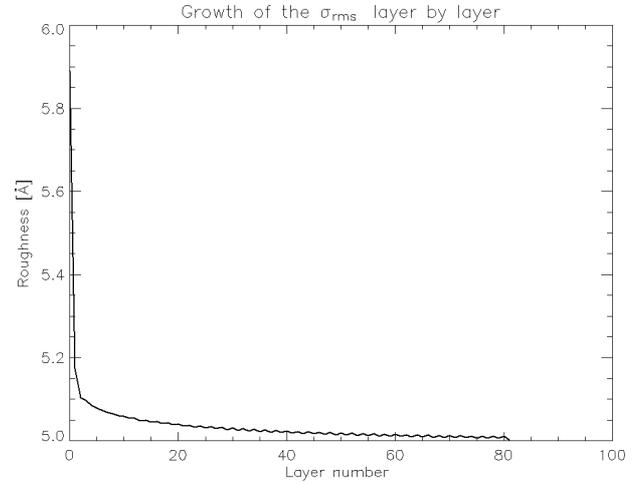

Figure 6: (left) Internal PSDs simulated with MPES for the multilayer sample D from substrate (dotted line) to final surface (black continuum line).
(right) Microroughness evolution for the multilayer sample D calculated from the internal PSD.

Table 9: Measured microroughness values for both the electroformed Nickel replicated substrate and the multilayer coating.

| $\sigma_{sub}$ [Å] | $\sigma_{ML}$ [Å] | $\sigma_{XRR}$ [Å] |
|---|---|---|
| 5.4 | 4.8 | 5 |

Table 10: Growth parameters used to simulate the microroughness evolution of the multilayer coating for the sample D.

|  | $\Omega$ | $l^*$ [nm] | n |
|---|---|---|---|
| Mo layers | $50 \cdot \Omega_{0,Mo}$ | 7.8 | 3 |
| Si layers | $1 \cdot \Omega_{0,Si}$ | 3.2 | 3 |

**Sample E**

For this sample the PSD was measured only at wavelengths smaller than 10 μm. This W/Si multilayer coating is formed by two stacks of bi-layers, with 7 thicker outer bi-layers (constant d = 130 Å) followed by 40 deeper ones with 38 Å d-spacing (see table 1). The deposition facility[15] used to coat the sample allowed to limit the microroughness evolution at a very low level ($\sigma_{ML}$ = 1.95 Å): more precisely, the growth of microroughness amounts to only 1 Å rms with respect to the initial substrate level (see tab. 11). The found parameters values are listed in tab. 12. Fig. 7 (left), shows the result of the simulated PSD evolution, the growth is well fitted in the measurement spectral range. In fig. 7 (right) we show the evolution of roughness rms: the trend is broken in correspondence to the change of thickness of the layers at the 40[th] bi-layer, since the evolution of the microroughness is strongly dependent on the thickness $\tau$ of each layer.

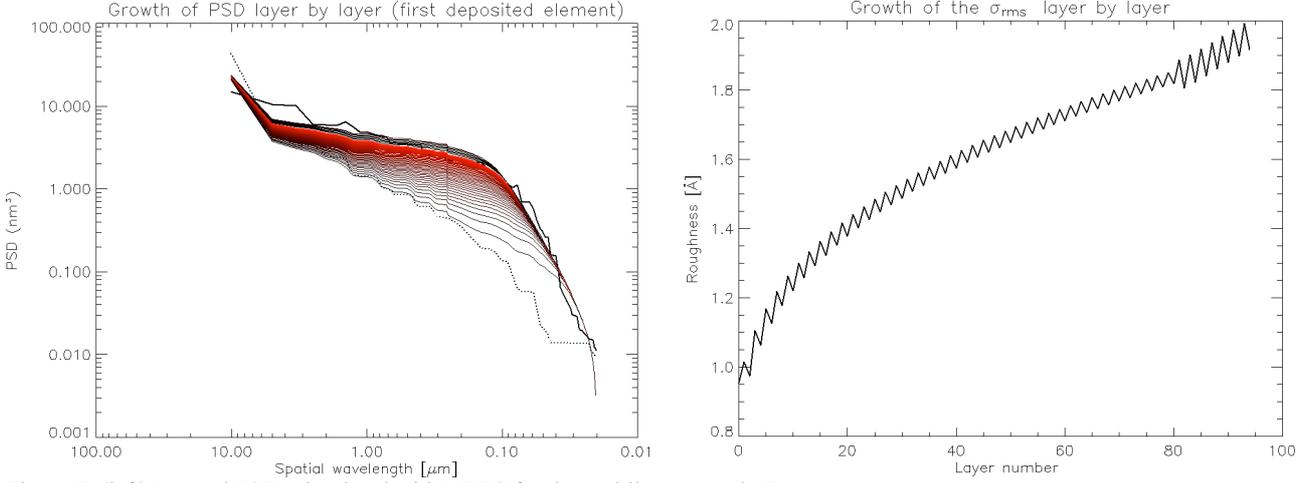

Figure 7: (left) Internal PSDs simulated with MPES for the multilayer sample E.
(right) Microroughness evolution for the multilayer sample E computed from the internal PSD. The slope change occurs at the transition between the two stacks h two different d-spacings.

Table 11: Measured roughness values for the multilayer sample E. Notice the very good quality of the used fused silica substrate.

| $\sigma_{sub}$ [Å] | $\sigma_{ML}$ [Å] | $\sigma_{XRR}$ [Å] |
|---|---|---|
| 0.97 | 1.95 | 3 |

Table 12: Growth parameters adopted to simulate the microroughness growth.

|  | $\Omega$ | $l^*$ [nm] | n |
|---|---|---|---|
| W layers | 43 · $\Omega_{0,W}$ | 4.8 | 4 |
| Si layers | 1 · $\Omega_{0,Si}$ | 3.3 | 4 |

## 5. RESULTS VERIFICATION BY X-RAY SCATTERING MEASUREMENTS

The PSD evolution in analyzed multilayer stacks has received an independent confirmation from of X-ray Scattering (XRS) technique. XRS is a powerful microroughness characterization tool for X-ray reflecting surfaces because the scattered intensity is proportional to the surface Power Spectral Density, hence a XRS measurement is able to return the surface PSD averaged over the irradiated area. This well-known relation has been extensively used by us to measure the finishing levels and the spectral properties of substrates for X-rays and neutron mirrors[17,22]. For multilayers, XRS is determined by the interference of scattered rays at each interface of the multilayer stack, therefore this technique allows an in-depth characterization of interfacial microroughness. In particular, the set of internal PSD we calculated from topographical data using MPES can be used also in order to compute the XRS diagram expected from the sample reflection at a definite incidence angle (e.g. when X-rays strike on the sample at the 1[st] Bragg peak). The agreement with the experimental scans, obtained at INAF/OAB using a Bede-D1 diffractometer, is an independent confirmation of the PSD evolution and of the correctness of the assumed modelization.

The relation PSD-XRS for single boundaries which microroughness is described by a single PSD is expressed by a well-known formula[21,23]:

$$\frac{1}{I_0}\frac{dI_s}{d\vartheta_s} = \frac{16\pi^2}{\lambda^3}\sin\vartheta_i \sin^2\vartheta_s R_{i_s} P(f) \qquad (7)$$

where $P(f)$ is the surface 1D Power Spectral Density as a function of spatial frequency $f$, $I_0$ is the incident flux, $I_s$ represents the scattered power at the scattering angle $\vartheta_s$, $\vartheta_i$ is the (grazing) incidence angle, $\lambda$ the wavelength of incident X-rays, and the *polarization factor* $R_{is}$ is related to the zero-roughness surface reflectivity $r$ at both angles of incidence and scattering, according to the *Rayleigh-Rice* theory[21]:

$$R_{is} = [r(\vartheta_i)r(\vartheta_s)]^{1/2} \tag{8}$$

the spatial frequency $f$ (or the reciprocal spatial wavelength $l$) is simply related to the radiation wavelength and to the incidence and scattering angles by the known grating formula:

$$l = \frac{\lambda}{\cos\vartheta_i - \cos\vartheta_s} \tag{9}$$

In a multilayer coating the equation (7) cannot be applied because several interfaces are generally involved in the reflection/scattering process: since the PSD evolves throughout the stack and because of (partial) correlation of roughness profiles, the resulting scattering pattern will not be, in general, a simple superposition of scattering diagrams of each multilayer interface, but will exhibit interference signatures. A very complete treatment of X-ray scattering in multilayers is developed in the framework of a rigorous formalism[24]. For our purposes – the verification of MPES results correctness – we have used a generalization of the eq. (7):

$$\frac{1}{I_0}\frac{dI_s}{d\vartheta_s} = \frac{16\pi^2}{\lambda^3}\sin\vartheta_i \sin^2\vartheta_s R_{is}\left[\sum_{j=0}^{N} T_j^2 P_j + 2\sum_{j>m}(-1)^{j+m} T_j T_m C_{jm}\cos(Q_z\Delta_{jm})\right] \tag{10}$$

The XRS diagram from a multilayer had already been derived by Kozhevnikov in a similar form[25]. The eq. (10) had been derived by one of us (D. Spiga) following a completely different approach[26]: as we will soon see, this equation is very suitable to describe the XRS diagram from the MPES outputs.

In this formula, $P_n$ ($j = 0, 1… N$) are the interfacial PSDs of the multilayer, $T_j$ indicates the electric field amplitude in the $j^{th}$ layer, normalized to the external incident electric field, which intensity is expressed by $I_0$: the sum is extended to all the multilayer interfaces. The polarization factor $R_{is}$ has the same expression as in eq. (8), whereas $r$ is now the single boundary reflectivity of the couple spacer/absorber of the multilayer.

The scattering diagram depends also on an interference term (the second sum in [ ] brackets), depending on the cross-correlations $C_{jm} = \text{Re}(\hat{z}_j^* \hat{z}_m)$ of all couples of interfaces. In this work all $\hat{z}_j^* \hat{z}_m$ are supposed to be already real (i.e., at each spatial frequency there is no phase shift between the spectral components of profiles while propagate trough the stack). $\Delta_{jm}$ is the distance between the $j^{th}$ and the $m^{th}$ interface, and $Q_z$ is the perpendicular component of scattering vector:

$$Q_z = \frac{2\pi}{\lambda}(\sin\vartheta_i + \sin\vartheta_s) \tag{11}$$

In eq. (10) $\lambda$ (= 1.541 Å) is known, $\vartheta_i$, $\vartheta_s$, $I_0$, $I_s$, can be are measured directly, $R_{is}$ is computed from Fresnel equations, the $T_j$ coefficients can be derived from the multilayer reflectivity, computed using a standard method[20,24]. In this work the incidence we simplify the $T_j$ coefficients by taking XRS detector scans with incidence angle at the $k^{th}$ Bragg peak: therefore, the electric field falls exponentially[24] in the stack. It can be proved, by means of simple calculations (D. Spiga[26]), that the electric field *amplitude* attenuation coefficient equals approximately $\xi = 2r^{1/2}\sin(\pi k\Gamma)$. For this result to hold, the single-boundary reflectivity *r has to be much less than 1*, and the electric field must be completely either reflected or absorbed before the end of the multilayer stack. In this work we always performed measurements at the 1st Bragg peak, and the mentioned conditions are fulfilled.

In order to verify the MPES results, we substituted in the eq. (10) the internal PSDs $P_j$ reported in the previous section. The cross-correlations can be, instead, calculated as follows: let us indicate with $z_m(x,y)$ the surface profile of $m^{th}$ interface in the stack. The $(m+1)^{th}$ interface partially replicates the profile: in terms of the profiles Fourier transforms,

$$\hat{z}_{m+1}(f) = a(f)\hat{z}_m(f) \tag{12}$$

where $a(f)$ is the replication factor of the $(m+1)^{th}$ layer. The profile term related to the single–layer PSD (eq. 4) is ignored because it is a random one, uncorrelated with the $m^{th}$ interface. The Fourier transforms are supposed to depend only on the radial frequency $f$ rather than on $f_x$ and $f_y$ independently because of the samples isotropy; so does the replication factor. Moreover, $a(f)$ is assumed to be real since we supposed that the multilayer was grown along the substrate normal, therefore the surface profiles are replicated with no lateral shift, and spectral components of the same frequency in adjacent layers should not be affected by phase changes. Iterating the previous equation up to the $j^{th}$ layer (we omit the dependence on the frequency for sake of clarity), we have for the correlated part of the $j^{th}$ profile,

$$\widehat{z}_j = a_j a_{j-1} \cdots a_{m+1} \widehat{z}_m \qquad (13)$$

and the $jm^{th}$ 2D cross-correlation simply results by taking the conjugate of (13) times $\widehat{z}_m$:

$$C_{jm} = \widehat{z}_j^* \widehat{z}_m = a_j a_{j-1} \cdots a_{m+1} \widehat{z}_m^* \widehat{z}_m = a_j a_{j-1} \cdots a_{m+1} P_m \qquad (14)$$

here $P_m$ denotes the bidimensional Power Spectral Density of the $m^{th}$ layer. All $C_{jm}$ cross-correlations should be converted[21] to the 1D form in order to be substituted in the eq. (10). The results of eq. (10) have been corrected for the absorption and the refraction of X-rays in the analyzed multilayer coatings; finally, the XRS diagram was degraded at the actual angular resolution of the detector (300 arcsec). The comparison of experimental XRS results for samples A, B, C, D (achieved with a BEDE-D1 diffractometer at 8.05 keV) with the predictions of eq. (10) is presented in figs. 7-10.

In order to highlight the PSD evolution in the multilayer we compared the simplified case of an hypothetical scattering diagram (left plots) computed by assuming a constant PSD, equal to that of the outer surface of the multilayer, and with a constant replication factor to compute $C_{jm}$. Even though the replication factor was chosen in order to achieve the best possible fit, we will see that the agreement of model with the results is unsatisfactory. Moreover, the replication factor value was chosen without any physical basis, therefore the calculation has a large degree of arbitrariness.

On the contrary, the agreement of XRS data with the predictions of the more exact findings of MPES (i. e., accounting for the PSD evolution throughout the stack), along with eq. (10), is *very satisfactory*: moreover, we did not need any arbitrary setting of replication factors, since they are already computed from the growth parameters that fit the PSD growth. This is a confirmation of the correctness of modelization for microroughness growth in the considered multilayer samples.

**Sample A**

The comparison of measured and simulated X-ray scattering at the 1$^{st}$ Bragg peak incidence angle (3925 arcsec) is presented in fig. 8. Since the XRS experimental scan was taken several months after the deposition, we could observe a relevant XRR peak reduction, that can be ascribed to interdiffusion. Assuming that the external microroughness remained unchanged (8.2 Å) with respect to the measured value soon after the deposition, the $\sigma_{diff}$ value for diffusion inferred from XRR fit is 9.1Å: in order to account for interdiffusion in computing the XRS diagrams, we lowered the single-boundary reflectance value $r$ (see eq. 10) by a Debye-Waller factor computed from $\sigma_{diff}$. If we assume all PSD to be equal in the stack and a constant correlation degree, we obtain the result in fig. 8 (left): the height of peaks are clearly overestimated. The PSD evolution of fig. 3 was instead adopted to compute the scattering diagram in fig. 8 (right); the shape of the first XRS peak is fitted much better, and the second peak almost disappeared, as in the experimental curve. The Yoneda peak is overestimated in the simulation due to the imperfect modelization of the critical angle in the function $r(\vartheta)$ (see the eq. 8).

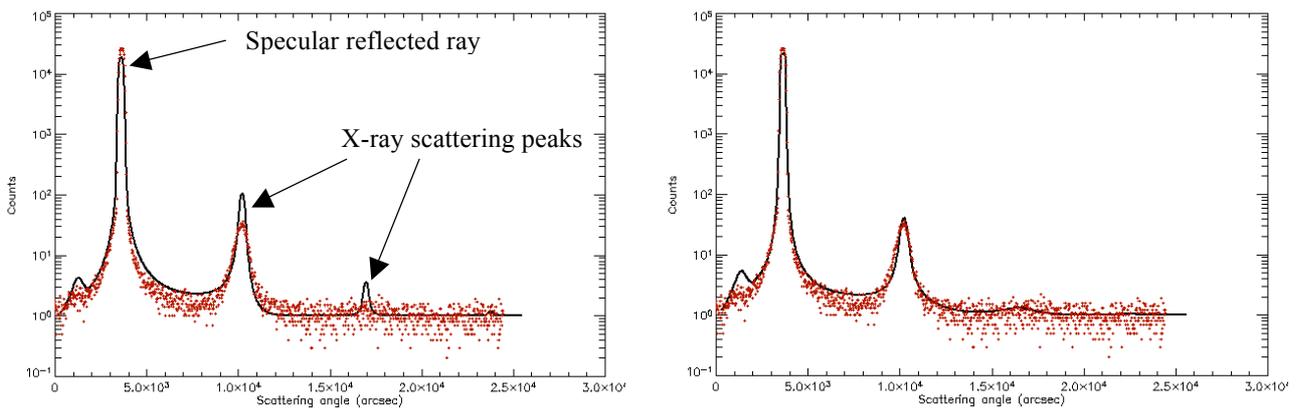

Figure 8: X-ray scattering diagram for the sample A, $\vartheta_i = 3935$". The modelization (solid line) is compared with the experiment (dots). *(left)* without PSD evolution. *(right)* using the PSDs and the cross-correlations computed from the growth analysis with MPES. The improvement is apparent.

**Sample B**

For this sample the XRS was measured only at angles larger than that of specular reflection: a $\sigma_{diff}$ rms interdiffusion of 9.5 Å, also inferred from XRR fit, was used in modelling the XRS diagrams. The adopted d-spacing (52.5 Å) and the Γ factor (0.365) are consistent with the d-spacing instability inferred from the XRR analysis. The comparison of measured and simulated X-ray scattering at the 1$^{st}$ Bragg peak incidence angle (3320 arcsec) is plotted in fig. 9. The height of peaks is better fitted by the PSD trend computed with MPES than the simplified model without PSD evolution. However, this is less evident than with the sample A, due to the weaker PSD change throughout the stack.

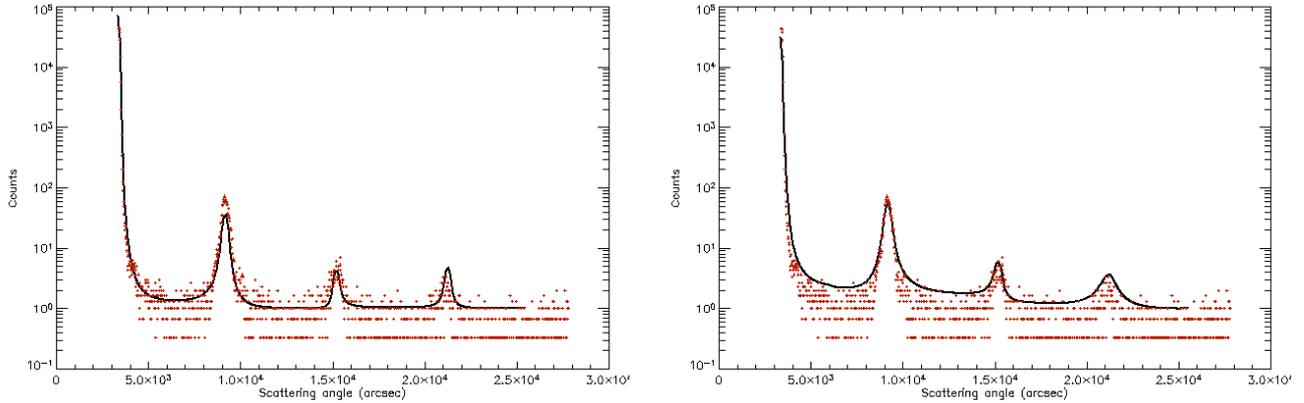

Figure 9: X-ray scattering diagram for the sample B. The modelization (solid line) is compared with the experiment (dots). *(left)* without PSD evolution. *(right)* using the PSDs and the cross-correlations computed from the growth analysis with MPES. The improvement is apparent.

**Sample C**

The simulated X-ray scattering diagram at the 1$^{st}$ Bragg peak incidence angle (3083 arcsec) is compared to the simulation in fig. 10. The PSD evolution fits very well the experimental XRS curve along with the eq. (10): the improvement with respect to the XRS diagram computed from a single PSD is apparent. Not only the heights of peaks could be well fitted, but also their shape are very well simulated. Even though the modelization was performed over a limited spatial frequency (10 – 0.02 μm), all the XRS curve could be simulated because all larger wavelengths scatter at angles smaller than the angular resolution of the measurement. Therefore, such scattered rays cannot be distinguished from the specularly-reflected ray.

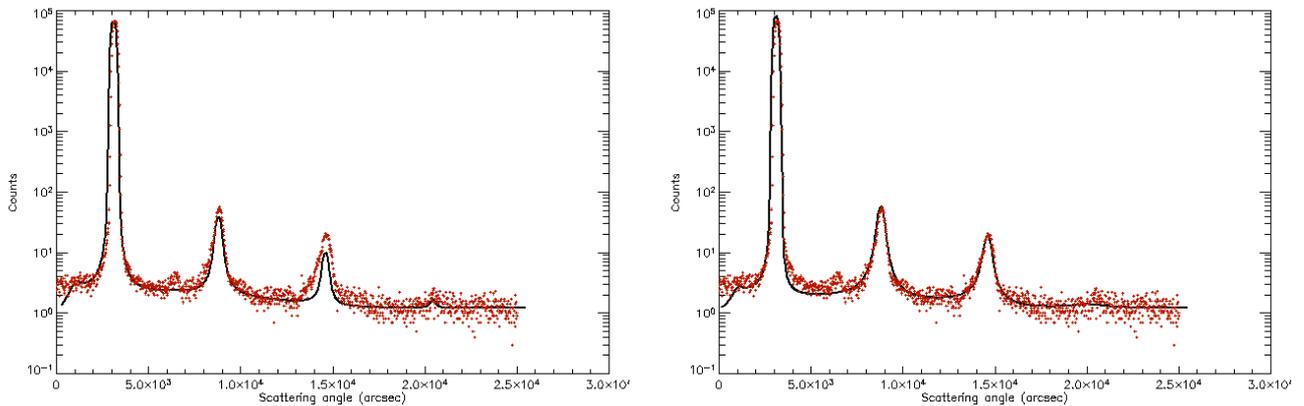

Figure 10: X-ray scattering diagram for the sample C. The modelization (solid line) is compared with the experiment (dots). *(left)* without PSD evolution. *(right)* using the PSDs and the cross-correlations computed from the growth analysis with MPES. The improvement is apparent.

**Sample D**

The comparison of measured and simulated X-ray scattering at the 1st Bragg peak incidence angle (2484 arcsec) is shown in fig. 11. The evidence of PSD evolution results clearly from the comparison of the two plots: the fit is clearly improved by computing the XRS diagram with the PSD decreasing trend shown in fig. 6. The height and the shape of XRS peaks match much better the XRS experimental data. Notice in particular how the 3rd scattering peak is broader and lower when we account for the PSD evolution, due to the diminishing correlation degree at large spatial frequencies. This is also a solid confirmation of the PSD improvement at large frequencies, already explained with MPES.

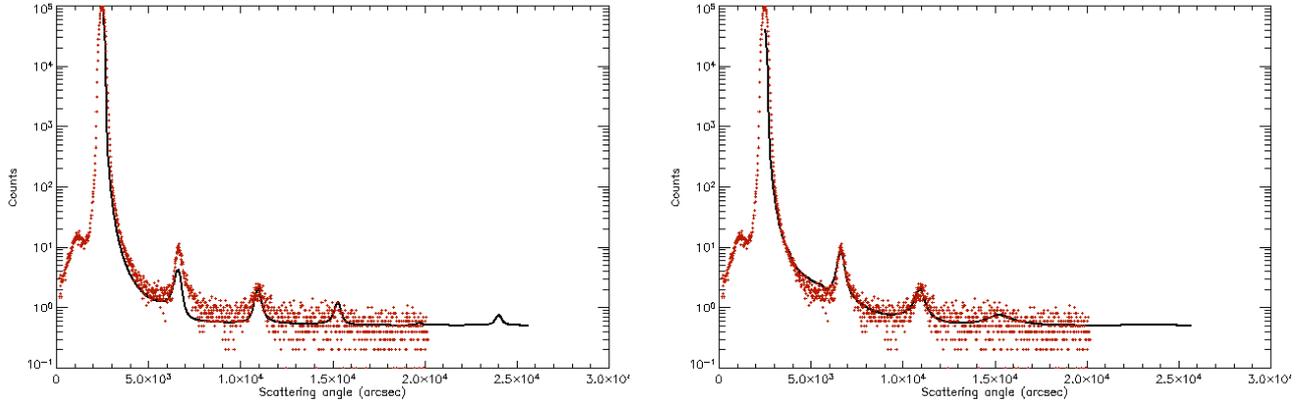

Figure 11: X-ray scattering diagram for the sample D. The modelization (solid line) is compared with the experiment (dots). *(left)* without PSD evolution. *(right)* using the PSDs and the cross-correlations computed from the growth analysis with MPES. The improvement is apparent. A 3 Å rms of interdiffusion were assumed in the simulation.

## 5. CONCLUSIONS

The exposed results showed, for a wide set of multilayer coatings, that an analysis on the microroughness PSD evolution is possible over a wide spectral range of spatial wavelengths in the framework of the formalism developed by Stearns[5]. We proved these results by implementing this model in the MPES program, providing it with a useful Graphical User Interface that makes easier the search for the growth parameters that characterize the thin film growth. All PSD evolution from the substrate up to the outer multilayer surface could be explained in terms of these few parameters, as well as the correlation between interfaces. The model/program apparently works well for either roughening and smoothing of surface profiles.

The correctness of the model was also verified by means of an independent tool (the X-ray Scattering) that is able to probe in-depth the microroughness PSD over a wide area fraction of the sample. Therefore, this last method also takes the advantage of a great statistical significance, and allowed us to highlight the usefulness of a tool like MPES in the prediction of the X-Ray scattering pattern from a multilayer coatings, since the X-ray simulated diagrams calculated from the inferred PSD evolution are in very satisfactory agreement with the experimental ones.

Future developments of this work will be aimed to the extension of MPES to the prediction of the PSD evolution in graded multilayers for optics of hard and soft X-ray telescopes: the results will also be checked by means of X-ray scattering measurements. After the verification of these results, MPES will be an important tool in the diagnostic of multilayer coatings in the framework of the optics development for future hard X-ray telescopes like SIMBOL-X.

## ACKNOWLEDGEMENTS

The authors are grateful to G. Nocerino, G. Valsecchi, G. Grisoni, M. Cassanelli (*Media-Lario techn.,* Bosisio Parini, Italy) for providing us with several W/Si multilayer samples and the sample D substrate. T. Maccacaro (*INAF/ Osservatorio Astronomico di Brera*, Milano, Italy) is also acknowledged to support this work. We thank V. Rigato (*Istituto Nazionale di Fisica Nucleare, Legnaro, Italy*) for the deposition of the Mo/Si multilayer sample and S. Romaine, P. Gorenstein, R. Bruni (*Harvard-Smithsonian Institute for Astrophysics, Boston, USA*), for depositing the W/Si sample E. R. Valtolina, S. Cantù (*INAF/ Osservatorio Astronomico di Brera*, Milano, Italy) are also acknowledged to execute the AFM and WYKO measurements. D. Spiga is indebted with MIUR (the Italian Ministry for University and Research) for the grant that made possible his fellowship.